# Frequency-Dynamics-Aware Economic Dispatch with Optimal Grid-Forming Inverter Allocation and Reserved Power Headroom

Fan Jiang, *Student Member, IEEE* and Xingpeng Li, *Senior Member, IEEE*

*Abstract*—The high penetration of inverter-based resources (IBRs) reduces system inertia, leading to frequency stability concerns, especially during synchronous generator (SG) outages. To maintain frequency dynamics within secure limits while ensuring economic efficiency, frequency-constrained optimal power flow (FCOPF) is employed. However, existing studies either neglect the frequency support capability and allocation of grid-forming (GFM) IBRs or suffer from limited accuracy in representing frequency dynamics due to model simplifications. To address this issue, this paper proposes a deep learning (DL)-based FCOPF (DL-FCOPF) framework. A DL model is first developed as a predictor to accurately estimate frequency-related metrics: the required reserved headroom and allocation of GFM IBRs, the rate of change of frequency and frequency nadir. After being trained with data obtained from electromagnetic transient simulations, the DL model is reformulated and incorporated into FCOPF. Case studies conducted on two test systems demonstrate the effectiveness of the proposed approach. Compared with the traditional OPF and linear FCOPF benchmarks, the DL-FCOPF can optimally coordinate SGs and IBRs with minimum cost, achieving desired frequency response, within an acceptable computing time. Furthermore, sensitivity analyses are conducted to identify the most suitable structure and linearization approach of the DL-based frequency predictor.

## Nomenclature

*Abbreviation*
| | |
|---|---|
| DL | Deep learning model. |
| FN | Frequency nadir. |
| GFM | Grid-forming control. |
| GFL | Grid-following control. |
| IBR | Inverter-based resource. |
| RoCoF | Rate of change of frequency. |

*Sets*
| | |
|---|---|
| $G$ | Set of synchronous generators. |
| $K$ | Set of transmission lines. |
| $I$ | Set of inverter-based resources. |
| $B$ | Set of buses. |

*Parameters*
| | |
|---|---|
| $f_{COI}$ | System frequency based on center of inertia. |
| $f_g$ | Local frequency at generator $g$. |
| $f_0$ | Base frequency. |
| $H_g$ | Inertia constant of generator $g$. |
| $a_g, b_g, c_g$ | Cost coefficients of generator $g$. |
| $P_b^{load}$ | Active load at bus $b$. |
| $P_{g,min}$ | Minimize active power limit of generator $g$. |
| $P_{g,max}$ | Maximum active power limit of generator $g$. |
| $P_{k,thm}$ | Thermal limit of line $k$. |
| $P_{IBR,i}^{max}$ | Maximum available active power of IBR $i$. |
| $P_{ref}$ | Base active power. |
| $P_{otg}$ | Active power output of outage generator. |
| $x_k$ | Reactance of line $k$. |
| $R_{lmt}$ | RoCoF limit. |
| $f_{lmt}$ | FN limit. |

*Variables*
| | |
|---|---|
| $P_{k,b}^{tbus}$ | Active power flow of line $k$ to bus $b$. |
| $P_{k,b}^{fbus}$ | Active power flow of line $k$ from bus $b$. |
| $P_{g,b}$ | Active power of generator $g$ at bus $b$. |
| $P_k$ | Active power flow of line $k$. |
| $\theta_p$ | Phase angle at bus $p$. |
| $\theta_q$ | Phase angle at bus $q$. |
| $R$ | Worst RoCoF during frequency response process. |
| $P_{IBR}$ | Active power of IBR. |
| $P_{GFM,i}$ | GFM active power output of IBR $i$. |
| $P_{GFL,i}$ | GFL active power output of IBR $i$. |
| $R^{DL}$ | RoCoF obtained from DL. |
| $f_{ndr}^{DL}$ | FN obtained from DL. |
| $P_{hdrm,i}^{DL}$ | Reserved active power of IBR $i$ obtained from DL. |
| $\alpha_i^{DL}$ | Allocation of GFM and GFL of IBR $i$ obtained from DL. |
| $\alpha_b$ | Allocation of GFM and GFL of IBR at bus $b$. |

## I. Introduction

With the global push toward decarbonization to prevent the depletion of fossil fuels and reduce carbon dioxide emissions, inverter-based resources (IBRs) have been widely deployed in power systems to facilitate the integration of renewable energy, particularly solar and wind power. In the United States, for instance, renewable energy accounted for approximately 21.4% of total electricity generation in 2023, compared to 12% in 1990 [1]. This transition from fossil-fuel-based traditional synchronous generators (SGs) to IBRs introduces several challenges. A primary concern is reduced grid inertia, which directly affects frequency stability [2].

According to the IEEE and CIGRE, frequency stability is the grid's ability to maintain steady frequency following a large power imbalance event [3]. When a disturbance occurs, such as a generator outage, the resulting power imbalance leads to a frequency decline. If the rate of change of frequency (RoCoF) becomes excessively high or the frequency nadir (FN) falls too low, under-frequency load shedding (UFLS) schemes would be triggered to disconnect loads [4], which may cause severe blackout events. Many power system incidents worldwide have been attributed to low system inertia and insuffi-

Fan Jiang and Xingpeng Li are with the Department of Electrical and Computer Engineering, University of Houston, Houston, TX, 77204, USA. (e-mails: fjiang6@uh.edu; xli83@central.uh.edu).

cient frequency response capability [5]-[7]. For example, a major blackout happened in Great British in August 2019, where low inertia was identified as the key factor exacerbating the disturbance and resulted in the loss of 1GW of electricity demand [7]. As renewable IBRs are still growing significantly, frequency stability issues would become more crucial, highlighting the need to ensure adequate frequency response to prevent cascading blackouts.

Research on improving frequency response performance has primarily focused on two aspects. The first set of methods improves IBR control. Most IBRs are controlled as grid-following (GFL) inverters, which inherently lack the capability to enhance system inertia [8], causing the rapid system frequency response that poses risks to system frequency stability. To address this limitation, grid-forming (GFM) control strategies have been developed, including droop control, virtual synchronous machine, and virtual oscillator control [9]-[10]. These technologies have been shown to effectively enhance system inertia and improve overall frequency response [12].

The second set of methods for enhancing frequency stability focuses on the system level. Since the frequency response performance is highly dependent on system conditions determined by economic dispatch, frequency dynamics can be constrained within grid operation models by adjusting the outputs of SGs and IBRs to enhance system frequency performance under contingencies. Several studies have incorporated frequency constraints into optimal power flow (OPF) and unit commitment (UC) models [13]-[19]. The main challenge in this process lies in balancing the accuracy of frequency response calculation with the computational burden. To meet the computational efficiency requirements of these optimization frameworks, the calculation of frequency-related metrics is obtained from the simplified system frequency response model derived from differential-algebraic equations (DAE) [20]-[21]. Many studies derive RoCoF from the swing equation of SGs, assuming that the worst RoCoF occurs at the onset of a contingency, while FN is often obtained from a simplified uniform system frequency dynamic model based on SG governor control [13]-[16]. Some work further considers local RoCoF that could be more severe than the system-wide RoCoF in certain areas [17], while others constrain FN indirectly through SG reserve requirements [18]-[20]. Although these heuristic methods significantly improve computational efficiency, they often lack accuracy because of simplifications.

Deep learning (DL) [22] has recently been applied to balance the trade-off between computational efficiency and accuracy in frequency response calculation, especially when considering the frequency support capability of GFM IBRs. [23] applies a DL model to predict frequency metrics under SG contingencies in frequency-constrained OPF (FCOPF), demonstrating higher accuracy than traditional approaches. [24] considers high IBRs penetration and employs a DL model to predict the FN for the UC problem, instead of using the low-order uniform system frequency dynamic model, achieving improved performance. [25] proposes a sparse neural network-based frequency-constrained UC model that substantially accelerates the solving speed though the application of sparse technologies. However, [23]-[25] neglect the frequency support capability of GFM IBRs. [26] incorporates the frequency support capability and reserved headroom of GFM IBRs, into a DL-based OPF model to schedule the virtual inertia. However, the optimal allocation between GFM and GFL IBRs has not been systematically investigated in [26], although such coordination is crucial since they are expected to coexist in future power systems [27]. To allocate GFM and GFL IBRs, [28] provides useful insights from a system strength perspective but does not account for economic aspects. Meanwhile, [29] examines the frequency support capability of wind turbines (WTs) and determines the GFM-GFL allocation of WTs at each bus. Nevertheless, [29] applies the linearized RoCoF calculation approach and indirect FN constraint to capture frequency dynamics, which means frequency metrics and the reserved headroom of WTs still suffer from challenges in accuracy.

To summarize, existing studies either neglect the frequency support and allocation of GFM IBRs, fail to jointly consider accuracy and computational efficiency, or overlook balancing the frequency stability with economic performance. To fill this gap, this paper proposes a DL-based FCOPF (DL-FCOPF) framework. The main contributions of this paper are summarized as follows:

- Frequency dynamics are incorporated into the FCOPF framework. Unlike the simplified frequency response model in [13]-[16], this paper employs a DL model as the frequency predictor to capture frequency dynamics. It is trained with data samples generated from electromagnetic transient (EMT) simulations, thereby accounting for detailed transient responses under contingencies. This enables the proposed DL-FCOPF to achieve higher accuracy compared to other simplified approaches.
- The frequency support capability of GFM IBRs is considered. The allocation of GFM and GFL IBRs and their reserved active power are optimized in the proposed DL-FCOPF framework, enforcing RoCoF and FN requirements with minimum cost. It can substantially enhance both the accuracy of frequency dynamics and overall system economic efficiency.
- The accuracy and the efficiency of the proposed model are verified. Comparative studies with two benchmark models, traditional OPF (T-OPF) without any frequency constraints and linear FCOPF (L-FCOPF) with linearized RoCoF constraint, are conducted on modified WSCC 9-bus and IEEE 39-bus systems for validation.
- The most suitable structure of the DL model and its linearization approach are identified through sensitivity analyses. Based on the modified IEEE 39-bus system, the performance of DL-FCOPF frameworks with different hidden layers and neurons is compared to determine the optimal DL model configuration. In addition, four different linearization approaches of the DL model are applied and compared to identify the one that achieves the best performance.

## II. OVERVIEW OF THE PROPOSED DEEP LEARNING-BASED FREQUENCY-CONSTRAINED OPTIMAL POWER FLOW

To accurately incorporate frequency dynamics into FCOPF, the DL-FCOPF is proposed in this paper. The whole procedure of developing and validating the proposed framework is shown in the flowchart in Fig. 1.

The detailed procedures are demonstrated as follows:

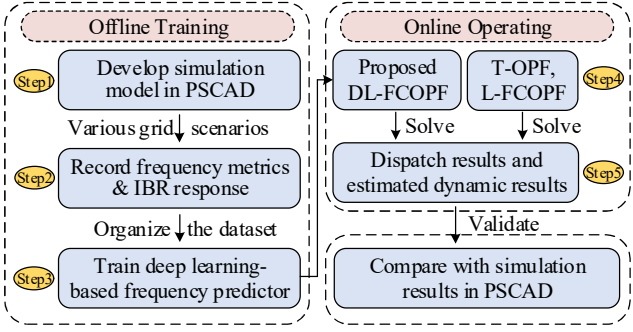

Fig. 1. Flowchart of the proposed DL-FCOPF framework.

*Step 1*: Develop an EMT simulation model in PSCAD to capture frequency response. This model generates the dataset for DL-based frequency predictor training in Step 2 and to validate the frequency dynamics of the dispatch results in Step 6. As a mature commercial platform, PSCAD is widely used for power system EMT simulations and dynamic performance analysis. To ensure the accuracy of the DL-based frequency predictor and DL-FCOPF framework, a PSCAD-based simulation model is developed in this step.

*Step 2*: Run the simulation model multiple times against various system scenarios, record frequency metrics and IBRs response. This step collects data samples for DL-based frequency predictor training, which plays a crucial role in determining the performance of the predictor. Variations in load levels, SG and IBR outputs, the GFM and GFL allocations, and contingency locations are considered for simulations. Specially, in each scenario, different load demands and active power outputs of SGs and IBRs are applied. Once the system reaches a steady state, a contingency event, assumed as a generator outage in this paper, is applied. The corresponding frequency response metrics are recorded and organized into the dataset. During this process, a dataset acquisition strategy is proposed to enhance the performance of the DL model, as detailed in Section III.B.

*Step 3*: Train the DL-based frequency predictor with obtained dataset. This step will develop a well-trained DL model for the DL-FCOPF framework. Different network structures, including various numbers of hidden layers and neurons, are tested and the configuration with the best performance is selected as the final model. The process for selecting the optimal structure is presented in the sensitivity analyses in Section V.C.

*Step 4*: Incorporate the trained DL model, reformulated as additional frequency stability constraints, into the proposed DL-FCOPF model. This step involves the linearization of activation function of the DL model and the incorporation of constraints of the DL model outputs to regulate frequency performance through adjustments in dispatch results. This step is critical for developing the DL-FCOPF model, as proper integration can significantly reduce the solving time while maintaining accuracy. Different linearization approaches are compared in the sensitivity analyses in Section V.C to identify the most efficient one.

*Step 5:* Solve the proposed DL-FCOPF model along with the benchmark models. In this step, two benchmark OPF models, T-OPF without frequency constraints and L-FCOPF with a linearized RoCoF constraint, are developed for comparison with the proposed model. All the models are solved under identical load and IBR levels for fair comparisons. The resulting dispatch outcomes will be applied to the EMT simulation model to verify the accuracy of the frequency dynamic response in the next step.

*Step 6*: Validate the economic dispatch results in the simulation model. This step aims to verify the accuracy of the frequency dynamic performance and assess the feasibility of the proposed DL-FCOPF framework. The dispatch results from the three OPF models under specific load and IBR levels are applied to the simulation model developed in Step 1. The simulated frequency dynamics are then compared to evaluate the accuracy and effectiveness of the proposed model.

Through the above steps, the proposed DL-FCOPF framework will be successfully developed and validated to ensure the accuracy and economic efficiency.

### III. DEEP LEARNING-BASED FREQUENCY PREDICTOR

To balance the trade-off between accuracy and efficiency of FCOPF while considering the frequency support capability and allocation of GFM IBRs, a DL model is employed as a frequency predictor trained on the data generated from EMT simulations. In this way, the predictor captures essential grid characteristics and fast IBR dynamics, and serves as its surrogate in OPF models. This section details the complete development of the DL-based frequency predictor, from data acquisition to training.

#### A. Structure of Deep Learning-based Frequency Predictor

The structure of the DL-based frequency predictor is illustrated in Fig. 2.

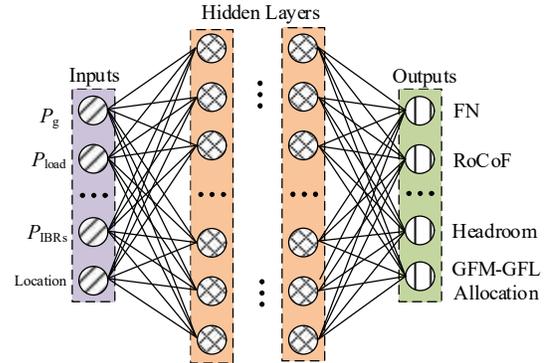

Fig. 2. Structure of DL-based frequency predictor.

The equation of this DL-based frequency predictor can be described by (1)-(4). The ReLU [32] is used as the activation function of this model.

$$\hat{z}_1 = W_1 z_0 + b_1 \quad (1)$$

$$\hat{z}_m = W_m z_{m-1} + b_m, 2 \leq m \leq NL - 1 \quad (2)$$

$$z_m = max(\hat{z}_m, 0), \ 1 \leq m \leq NL - 1 \quad (3)$$

$$F = W_{NL+1} z_{NL} + b_{NL+1} \quad (4)$$

where, $W_m$ and $b_m$ represent weight matrix and bias vector of layer $m$ respectively, $NL$ denotes the number of layers, $z_m$ indicates the preactivated values of neurons, $z_0$ stands for the inputs, and $F$ is the output vector.

The inputs to the predictor are extracted from system conditions, including SG outputs, load levels, the GFM and GFL IBR outputs, and the contingency location. The outputs correspond to frequency dynamic response related metrics that should be constrained in DL-FCOPF, namely GFM-GFL allocation ratio, reserved headroom of GFM IBRs (hereafter re-ferred to as headroom), RoCoF, and FN. These features collectively represent the frequency support capability of GFM IBRs, as explained below.

Typical GFL IBRs have limited response to the frequency variations, while GFM IBRs has the capability to provide fast frequency support. Fig. 3 illustrates the active power responses of GFM and GFL IBRs following a sudden SG outage.

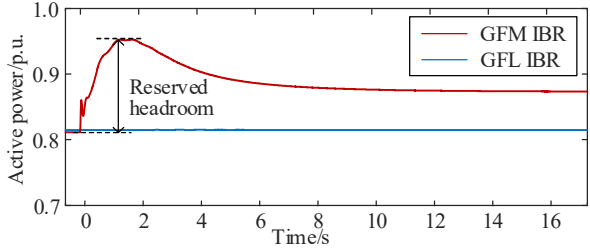

Fig. 3. Active power responses of GFM and GFL IBRs under SG outage.

It shows that the GFM IBR can respond rapidly to disturbances, providing fast frequency support and reaching its maximum active power output within 2s, thereby improving the overall frequency performance. GFM IBRs with different capacities can respond differently to frequency changes. Their maximum power deviation, referred to as the reserved headroom as illustrated in Fig. 3, reflects their frequency support capability, whereas it is difficult to determine precisely as it is a transient characteristic governed by DAEs. This challenge can be addressed by treating the GFM-GFL allocation and headroom as the outputs of the DL-based frequency predictor. In this way, the frequency support capability of the GFM IBRs can be predicted and incorporated into FCOPF.

### B. Training Data Acquisition Approach

The dataset quality has a critical impact on the performance and accuracy of DL models. In this paper, we ensure high-quality datasets by generating representable grid conditions and using high-fidelity EMT simulations for data generation.

Many studies rely on phasor analysis-based simulation platforms to generate frequency response datasets; however, these platforms may not fully capture system dynamics. In contrast, EMT-based simulation platforms such as PSCAD offer more accurate representations as they can capture fast IBR dynamics. Therefore, to ensure the reliability of the data source, this paper employs EMT simulation models in PSCAD to generate the dataset.

Once the simulation model is developed, the dataset is obtained from a series of independent runs with different operating conditions. In this paper, variations in load levels, the available power of IBRs, SG setpoints and contingency location are considered. The contingency events are generator outages. Since DL is data-hungry, a large number of EMT simulations are required to generate enough samples for DL model training. In a multi-generator system, capturing frequency response characteristics requires accounting for changes in each individual component, which cause the number of simulations to grow exponentially. This leads to unnecessary long simulation time, as large-scale EMT simulation are computationally expensive.

To address this issue, we adopt a strategy to reduce dataset size while maintaining the effectiveness of the trained DL model. First, one SG that is prone to tripping is selected, and its active power output is restricted to a specified value. The load level and IBR level are then varied, and for each combination of load and IBR levels, the T-OPF is solved to obtain the outputs of all SGs. These SG outputs, together with the restricted SG, load and IBR levels constitute one scenario. This process repeats to create various scenarios; then, EMT simulations are conducted under these scenarios, and the frequency dynamic responses are recorded. Next, the initial generation of the outage SG is changed, and the previous process is repeated to obtain the frequency dynamic response under different outage sizes. Noting that the values of SGs, load and IBRs are randomly selected within specified ranges. This procedure is applied to each SG that is prone to tripping. In this way, the number of simulations is significantly reduced while preserving the essential information within the dataset of representable scenarios.

For each simulation, the GFM-GFL allocations at each bus and headroom of each GFM IBR are recorded, the RoCoF and FN are calculated. The RoCoF measurement window typically ranges from 5 to 10 cycles [30]; in this paper, we adopt 10 cycles (0.167s for a 60 Hz system). The worst measured values are taken as the RoCoF labels. It is worth noting that frequency responses differ across buses, we use center of inertia (COI) [31] to represent the system frequency and calculate RoCoF and FN. The equation of COI is shown in (5).

$$f_{COI} = \frac{\sum_{g=1}^{G} f_g H_g}{\sum_{g=1}^{G} H_g} \quad (5)$$

After obtaining RoCoF, FN, headroom and GFM-GFL allocation for all scenarios, the resulting dataset is used in the training process. The trained DL-based frequency predictor is then integrated into FCOPF as frequency-related constraints to ensure frequency performance under the OPF dispatch points.

## IV. OPTIMAL POWER FLOW FRAMEWORKS

This section introduces three OPF frameworks. First, a tradition OPF without frequency constraint and a linear FCOPF with linearized RoCoF constraints are presented. Based on these two benchmark models, the DL-FCOPF framework is proposed to enable accurate and efficient integration of grid frequency dynamics.

### A. Traditional Optimal Power Flow

The objective function of T-OPF is to minimize total cost, as shown in (6).

$$\min \sum_{g=1}^{G}(a_g P_g^2 + b_g P_g + c_g) \quad (6)$$

The T-OPF constraints include the nodal power balance equations, power flow equations, generator output constraints, and transmission line thermal limits, which are presented in (7)-(10), respectively.

$$\sum P_{k,b}^{fbus} + P_b^{load} = \sum P_{k,b}^{tbus} + \sum P_{g,b} + P_{IBR,b}, \forall b \in B \quad (7)$$

$$P_k = \frac{\theta_p - \theta_q}{x_k}, \ p,q \in B, \forall k \in K \quad (8)$$

$$P_{g,min} \leq P_g \leq P_{g,max}, \forall g \in G \quad (9)$$

$$-P_{k,thm} \leq P_k \leq P_{k,thm}, \forall k \in K \quad (10)$$

### B. Linear Frequency-Constrained Optimal Power Flow

As explained earlier, conventional mathematical approaches for frequency calculation rely on simplified frequency response models. The worst RoCoF (hereafter referred to as RoCoF) is assumed to occur at the onset of a SG outage and can thus be derived from the generator swing equations, as shown in (11) [13]-[16].

$$R = -\frac{f_0}{2\sum_{i=1}^{G} H_i P_B} P_{otg} \quad (11)$$

Then the RoCoF constraint is expressed as

$$-\frac{f_0}{2\sum_{i=1}^{G} H_i P_{ref}} P_{otg} \geq R_{lmt} \quad (12)$$

The calculation of FN with the simplified uniform system frequency dynamic model still involves nonlinear components in its equations, making it time-consuming and inconsistent with real-time dispatch requirements; therefore, it is not incorporated into the L-FCOPF model.

In summary, the L-FCOPF model extends the T-OPF by incorporating the additional linearized RoCoF constraint (12).

### C. Deep Learning-based Frequency-Constrained Optimal Power Flow

To accurately incorporate the frequency dynamic performance and frequency support capability of IBRs into FCOPF models, the DL-based frequency predictor shown in Fig. 2 is employed to capture the frequency response metrics and integrated into the proposed DL-FCOPF in this subsection.

The activation function (3) of the DL-based frequency predictor is nonlinear and needs to be linearized before being incorporated into the DL-FCOPF framework. The commonly used linearization method is the BigM piecewise linearization (BPWL) without approximation, which can accurately reflect the performance of the DL model. However, due to the increased computational complexity of DL-FCOPF, three approximate linearization approaches, convex triangle area relaxation (CTAR), penalized convex triangle area relaxation (P-CTAR), and penalized convex area relaxation (PCAR) are introduced to reduce the solving time while maintaining the DL model performance [33]. These four approaches are illustrated in Fig. 4, and their corresponding mathematical formulations are expressed as follows.

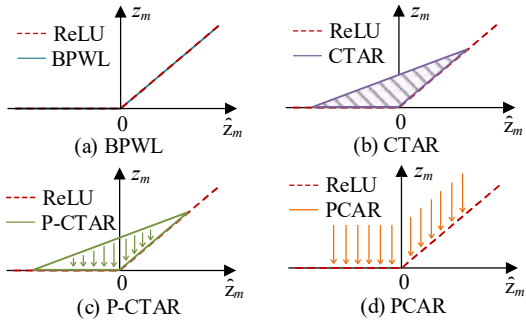

Fig. 4. Illustration of ReLU linearization approaches: a) BPWL, b) CTAR, c) P-CTAR, d) PCAR.

*1) BigM piecewise linearization:* This approach introduces binary variables to linearize the ReLU function as shown in Fig. 4(a). Consequently, the activation function (3) can be reformulated as (13a)-(13d).

$$z_m \leq \hat{z}_m - h_l(1 - B_m) \quad (13a)$$
$$z_m \geq \hat{z}_m \quad (13b)$$
$$z_m \leq h_u B_m \quad (13c)$$
$$z_m \geq 0 \quad (13d)$$

where $h_l$ and $h_u$ are the lower boundary and upper boundary of the value of all possible $\hat{z}_m$, and $B_m$ is a binary variable.

Although this approach avoids approximating the ReLU function, binary variables $B_m$ are introduced into the FCOPF model, which considerably decreases the solving efficiency.

*2) Convex triangle area relaxation:* The CTAR method constrains the feasible solution set to a triangular region, as shown in the purple area in Fig. 4(b); therefore, it involves approximation during the linearization. Its formulation is given by (13b), (13d) and (13e).

$$z_m \leq \frac{h_u}{h_u - h_l}\hat{z}_m - \frac{h_u \cdot h_l}{h_u - h_l} \quad (13e)$$

The CTAR method consists entirely of linear functions. By avoiding binary variables, it enhances the solving efficiency compared to BPWL.

*3) Penalized convex triangle area relaxation:* To reduce the approximation error of CTAR, the P-CTAR method introduces an additional penalty term in the objective function to enforce $Z_m$ to be close to the red dash line in Fig. 4(c) which represent the accurate ReLU function. It can be formulated by (13b), (13d), and (13e), with an additional term (13f) included in the objective function.

$$f^c = c_h \sum z_m \quad (13f)$$

where $c_h$ denotes the penalty coefficient.

*4) Penalized convex area relaxation:* This approach approximates ReLU without explicitly defining upper or lower boundaries, instead only using a penalty term to enforce $Z_m$ to approach the red dash line in Fig. 4(d). It can be formulated by (13b) and (13d), with (13f) as additional term in the objective function.

The performance of the four linearization approaches will be compared in Section V.C to identify the most effective one that achieves both high accuracy and solving efficiency.

After reformulating the nonlinear functions in the DL-based frequency predictor, constraints should be applied to its outputs to ensure they have the intended effect on the system. There are four kinds of outputs in the DL-based frequency predictor, which are RoCoF, FN, headroom and GFM-GFL allocation. RoCoF and FN can be constrained by (14)-(15).

$$R^{DL} \geq R_{lmt} \quad (14)$$
$$f_{ndr}^{DL} \geq f_{lmt} \quad (15)$$

The headroom of each IBRs can be constrained by (16).

$$P_{IBR,i} + P_{hdrm,i}^{DL} = P_{IBR,i}^{max}, \forall i \in I \quad (16)$$

This constraint ensures that the IBR outputs and their reserved headroom do not exceed their capability even during the transient response period. Typically, their capability corresponds to the active power output in maximum power tracking (MPPT) mode.

TABLE I
OBJECTIVES AND CONSTRAINTS OF T-OPF, L-FCOPF AND DL-FCOPF

| Model | Objective | Shared Constraints | Frequency Related Constraints |
|---|---|---|---|
| T-OPF |  |  | None |
| L-FCOPF | (6) | (7)-(10) | (12) |
| DL-FCOPF* |  |  | (1)-(2), (4), (13)-(18) |

* The constraints of (13) used in the DL-FCOPF model depend on the selection of DL linearization methods as explained above.

The allocation of GFM and GFL IBRs is restricted by constraints (17)-(18).

$$P_{GFM,i} = P_{IBR,i} \cdot \alpha_i^{DL}, \forall i \in I \quad (17)$$
$$P_{GFL,i} = P_{IBR,i} \cdot (1 - \alpha_i^{DL}), \forall i \in I \quad (18)$$

These two constraints determine the outputs of GFM and GFL IBRs based on their specified allocation.

In summary, the objectives and constraints of the three OPF models are presented in Table I.

## V. CASE STUDIES

In this section, the modified WSCC 9-bus and IEEE 39-bus systems are employed to evaluate the performance of the proposed DL-FCOPF framework. The detailed procedure is presented below.

### A. WSCC 9-bus system

A modified WSCC 9-bus system is adopted to evaluate the effectiveness of the proposed DL-FCOPF framework, whose structure is shown in Fig. 5. Compared with the standard system [34], the SG at bus 2 is replaced by GFM and GFL IBRs, while the SGs at bus 1 and bus 3 are divided into multiple identical units to ensure that the contingency capacity remains within a realistic range. The SGs connected to the same bus are identical, including governor and exciter control settings, inertia constants and cost coefficients.

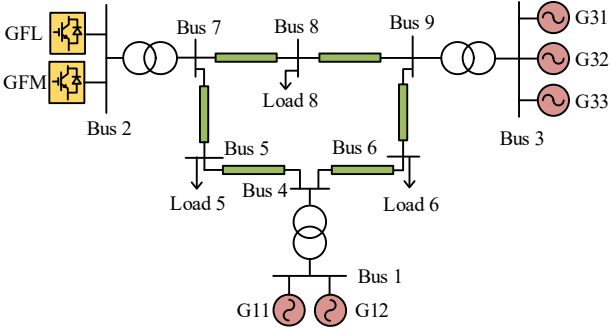

Fig. 5. Structure of the modified WSCC 9-bus system.

*1) Model development:* According the procedures shown Section II, the first step is to develop the modified WSCC 9-bus system EMT model in PSCAD, followed by the time-domain simulations under various scenarios - the associated frequency responses are also recorded. In these scenarios, the load level and the maximum available active power of IBRs vary from 90% to 110%, while the active power setpoints of the SGs at bus 1 and bus 3 range from 75% to 125%. The total IBR capacity is fixed, while the allocation of GFM and GFL IBR at bus 2 varies from 0 to 1. In each scenario, the contingency event is the sudden outage of a SG at bus 2 or bus 3. The locations of the contingencies are labeled in the dataset. The recorded frequency response metrices include RoCoF, FN, GFM IBR output power deviation, and GFM-GFL allocation ratio.

The active power output of each SG and IBR, the active loads at each bus, the contingency location labels, and the recorded frequency response metrics constitute the dataset for training the DL-based frequency predictor. A single-hidden-layer architecture with 32 neurons is adopted for this predictor, and its performance is shown in Fig. 6.

The two loss curves in Fig. 6(a) decrease as the epoch increases and then gradually stabilize at low loss values, indicating that the DL-based frequency predictor has been well-trained. The parity plots of RoCoF, FN and headroom in Fig. 6(b)-(d) further confirm that the DL model performs well on previously unseen data. This demonstrates the predictor is able to effectively capture frequency dynamics, thereby improving the accuracy of the proposed DL-FCOPF framework.

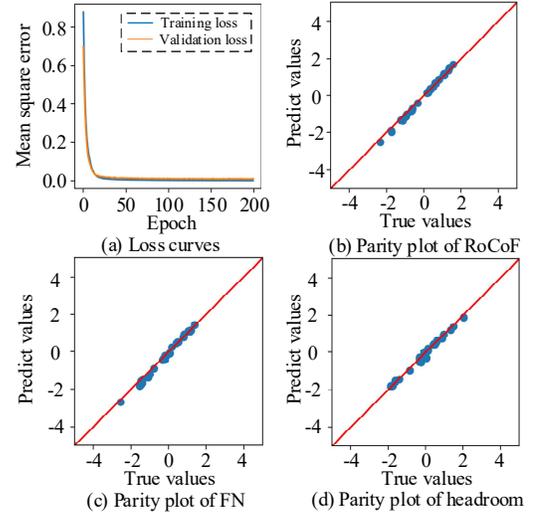

Fig. 6. Performance of the DL-based frequency predictor: (a) loss curves, parity plots of (b) RoCoF, (c) FN, (d) headroom.

Once trained, the DL-based frequency predictor is reformulated using the no-approximation linearization approach BPWL and embedded into the FCOPF framework, thereby enabling the dispatch model to account for frequency stability requirement in real time. In this paper, the RoCoF limit $R_{lmt}$ is set to -0.5Hz/s and the frequency nadir limit $f_{lmt}$ is specified as 59.5Hz [17]. The proposed DL-FCOPF, and two benchmark models, T-OPF and L-FCOPF, are implemented in Pyomo and solved on the same computer with a 12th Gen Intel® Core™ i7-12700 CPU @ 2.1 GHz and 32.0 GB of RAM.

*2) Results analysis:* To verify the efficiency of the proposed DL-FCOPF framework, all three OPF models are solved under the condition of 105% load level, 110% IBRs level, and a sudden loss of a SG at bus 3. The respective dispatch results are presented in Table II.

TABLE II
DISPATCH RESULTS OF OPF MODELS ON MODIFIED WSCC 9-BUS SYSTEM

| Models | $P_{IBR}$/MW | $\alpha_2$ | Total cost/$ | Solving time/s |
|---|---|---|---|---|
| T-OPF | 179.3 | N/A | 4323.83 | 0.18 |
| L-FCOPF | 179.3 | N/A | 4336.07 | 0.23 |
| DL-FCOPF | 165.77 | 69.9% | 4482.17 | 0.55 |

As shown in Table II, the maximum available active power of the IBR is 173MW. In the DL-FCOPF dispatch results, the IBR output is lower than the maximum available active power because a portion of the active power is reserved by the GFM IBR for frequency response support. This reserved power is compensated by SGs, which incur higher generator costs. Consequently, the total cost of the DL-FCOPF is higher than that of the other two models. The solving time is relatively high due to the binary variables introduced in the DL-based frequency predictor.

After solving the three OPF models, the dispatch results are implemented into the EMT simulation models in PSCAD respectively to verify the effectiveness of the proposed DL-FCOPF framework. The resulting frequency dynamic response curves are compared in Fig. 7. The FN, RoCoF and headroom obtained from the simulation model are regarded as the exact values and are compared in Table III with the estimated values obtained from the OPF models.

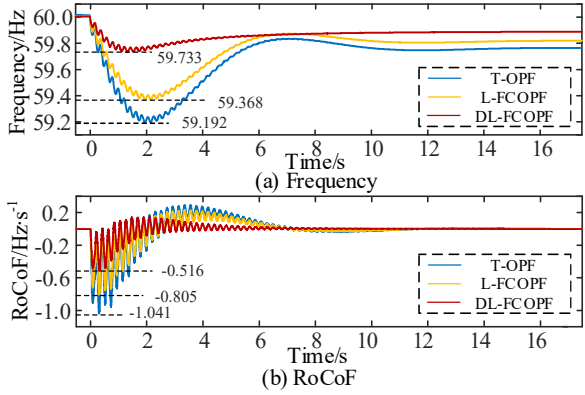

Fig. 7. Frequency dynamics on modified WSCC 9-bus system with dispatch results from T-OPF, L-FCOPF and DL-FCOPF respectively: (a) Frequency, (b) RoCoF.

TABLE III
OPF DISPATCH RESULTS AND DYNAMIC SIMULATION RESULTS

| Models | | FN | RoCoF | Headroom |
|---|---|---|---|---|
| T-OPF | Estimated | N/A | N/A | N/A |
| | Exact | 59.73 Hz | -1.041 Hz/s | 0 MW |
| | Error | N/A | N/A | N/A |
| L-FCOPF | Estimated | N/A | -0.5 Hz/s | N/A |
| | Exact | 59.32 Hz | -0.803 Hz/s | 0 MW |
| | Error | N/A | 37.73% | N/A |
| DL-FCOPF | Estimated | 59.71 Hz | -0.5 Hz/s | 13.53 MW |
| | Exact | 59.73 Hz | -0.516 Hz/s | 11.89 MW |
| | Error | 0.05% | 3.10% | 13.81% |

As shown in Fig. 7, the DL-FCOPF's dispatch result has the best frequency performance among the three models. The results from the T-OPF without any frequency constraints indicate that, under a contingency, both the FN and RoCoF exceed their thresholds, which may trigger UFLS. Although the RoCoF and FN obtained from the L-FCOPF are improved compared with those of the T-OPF, they still violate the respective thresholds, indicating that a linear RoCoF constraint provides only limited improvement. In contrast, the frequency response metrics from the DL-FCOPF indicate that the FN is effectively constrained within the threshold and RoCoF closely approaches the threshold, thereby ensuring frequency stability under a contingency.

In Table III, since no estimated value are available from T-OPF, only its exact values are reported. For the other two OPF models, though L-FCOPF applies the same RoCoF threshold as the DL-FCOPF, its error between the estimated and exact values reaches 37.73%, whereas for the DL-FCOPF is only 3.10%. In addition, the errors of FN and headroom are 0.05% and 13.81%, respectively, which are within acceptable ranges. These results demonstrate that the proposed DL-FCOPF accurately captures frequency dynamics before real-time and effectively enhances frequency stability in the look-ahead economic dispatch by proposition the system.

To further highlight the importance of incorporating GFM-GFL allocation, Table IV presents a comparison of dynamic results under different allocations.

TABLE IV
COMPARISON OF RESULTS UNDER DIFFERENT GFM-GFL ALLOCATIONS

| GFM Allocation Ratio | FN/Hz | RoCoF /Hz·s$^{-1}$ | $P_{hdrm}$/MW | Total Cost/$ |
|---|---|---|---|---|
| 50% | 59.686 | -0.573 | 9.995 | 4400.02 |
| 69.9% (DL-FCOPF) | 59.736 | -0.516 | 11.886 | 4414.44 |
| 90% | 59.769 | -0.470 | 13.314 | 4425.34 |

In Table IV, the optimal allocation of 69.9% GFM and 30.1% GFL IBRs is obtained from DL-FCOPF. Without explicit GFM-GFL allocation information from the DL-FCOPF framework, the allocation of GFM and GFL IBRs and the reserved headroom would deviate from these optimal values. As shown in Table IV, a lower allocation can result in insufficient frequency support, leading to degraded FN and RoCoF performance. Conversely, a higher GFM allocation may lead to an overly conservative solution, thereby reducing the grid operational efficiency with higher cost. Thus, it is critical for DL-FCOPF to determine the optimal allocation of GFM and GFL to improve frequency stability and economic efficiency.

In summary, the proposed DL-FCOPF framework more accurately captures frequency dynamics and IBRs frequency support capability than the other two OPF models, thereby effectively enhancing frequency stability.

### B. IEEE 39-bus system

The modified IEEE 39-bus system is tested to verify the scalability of the proposed DL-FCOPF model. Compared to the standard system [35], SGs at bus 36 and bus 39 are replaced by IBRs with both GFM and GFL control mechanisms available, as shown in Fig. 8. The contingency events are the outages of the largest SG in terms of power capacity.

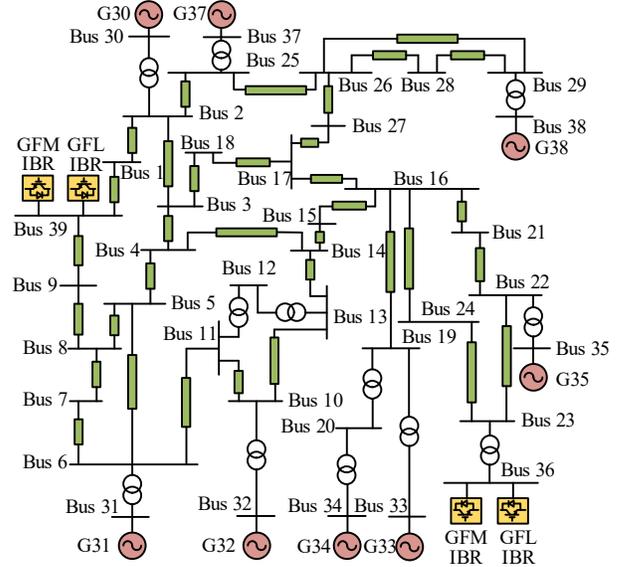

Fig. 8. Illustration of the modified IEEE 39-bus system.

*1) Model development:* Similar to the modified WSCC 9-bus system, a simulation model of the modified IEEE 39-bus system is developed in PSCAD. A series of simulations are conducted to generate the dataset for training the DL model. The load level varies from 70% to 110%, while the IBR level varies from 90% to 110%. The contingency event considered is the outage of the largest-capacity generator. After constructing the dataset, the DL-based frequency predictor with single hidden layer and 32 neurons is trained. It is worth noting that the outputs of the DL-based frequency predictor include two headroom values and two GFM-GFL allocation ratios, corresponding to the two IBRs located at two different buses respectively in the modified IEEE 39-bus system. After training, the DL-based frequency predictor is reformulated and incorporated into the FCOPF model.

*2) Results analysis:* The proposed DL-FCOPF and the two benchmark models T-OPF and L-FCOPF are solved under a grid condition of 105% load level, 115% and 110% generation levels for the two IBRs respectively. The contingency considered is the outage of the SG with the largest capacity, denoted as G38 in Fig. 8. This scenario is designed to assess the impact of the proposed DL-FCOPF on the dispatch of SGs. To reflect practical operational requirements, the IBRs are assumed to maintain a specified allocation of GFM and reserved active power for frequency support. In this process, the DL-FCOPF is first solved, and the resulting IBR outputs are subsequently applied in T-OPF and L-FCOPF for comparison. The dispatch results of the T-OPF, L-FCOPF and DL-FCOPF are summarized in Table V. The frequency dynamics from the simulation models based on these dispatch results are compared in Fig. 9. The exact values and estimated values from the OPF runs and dynamic simulation runs are presented in Table VI.

TABLE V
OPF DISPATCH RESULTS ON THE MODIFIED IEEE 39-BUS SYSTEM

| Models | G38/MW | Total cost/$ | Solving time/s |
|---|---|---|---|
| T-OPF | 859.5 | 351026 | 0.22 |
| L-FCOPF | 859.5 | 351026 | 0.26 |
| DL-FCOPF | 681.1 | 354518 | 1.03 |

TABLE VI
DYNAMIC VALIDATION RESULTS ON THE MODIFIED IEEE 39-BUS SYSTEM

| Models | | FN/Hz | RoCoF /Hz·s$^{-1}$ | $P_{\text{hdrm},36}$ /MW | $P_{\text{hdrm},39}$ /MW |
|---|---|---|---|---|---|
| T-OPF | Estimated | N/A | N/A | N/A | N/A |
| | Exact | 59.693 | -0.660 | 36.362 | 69.998 |
| | Error | N/A | N/A | N/A | N/A |
| L-FCOPF | Estimated | N/A | -0.314 | N/A | N/A |
| | Exact | 59.693 | -0.660 | 36.362 | 69.998 |
| | Error | N/A | 52.42% | N/A | N/A |
| DL-FCOPF | Estimated | 59.769 | -0.5 | 28.172 | 52.108 |
| | Exact | 59.768 | -0.501 | 27.330 | 51.711 |
| | Error | <0.01% | 0.20% | 3.08% | 0.77% |

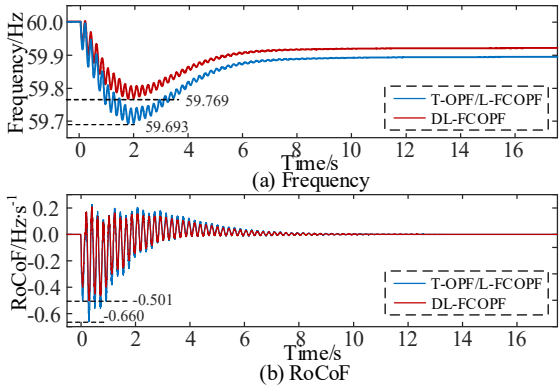

Fig. 9. Frequency dynamics in scenario 2 from simulations under dispatch results from T-OPF, L-FCOPF and DL-FCOPF: (a) Frequency, (b) RoCoF.

The results presented in these tables and figure indicate that the T-OPF and L-FCOPF yield identical dispatch and dynamic results because the estimated RoCoF from L-FCOPF is less than -0.5Hz/s, indicating that the linearized RoCoF constraint is inactive and will not affect the solution. This observation suggests that, in certain scenarios, the linearized RoCoF constraint fail to capture grid frequency dynamics, leading to noticeable inaccuracies in L-FCOPF results. Table V demonstrates that the DL-FCOPF yields distinct SGs dispatch outcomes, even under the same IBRs conditions as the other two OPF models. Furthermore, Table VI shows that the dynamic results of the DL-FCOPF solution exhibit negligible error when validated against the exact EMT simulation results from PSCAD, confirming the accuracy of the proposed DL-FCOPF.

*C. Sensitivity Analysis*

When integrating the DL model into optimization frameworks, the overall performance is influenced by several factors. The numbers of hidden layers and neurons affect both accuracy and computational efficiency, as it determines the performance of the original DL model, as well as the numbers of additional variables and constraints introduced into the FCOPF model. The linearization approaches applied to the activation function during the DL model reformulation also impacts the accuracy and solving speed of DL-FCOPF framework, since it introduces extra constraints and variables, and in some cases, even changes the optimization type (e.g., from linear programming to mixed-integer linear programming). To identify the most suitable DL model configuration and linearization approach for the FCOPF framework, a series of sensitivity analysis considering these factors are conducted on the modified IEEE 39-bus system in this subsection.

*1) Numbers of hidden layers and neurons:* The numbers of hidden layers and neurons determines how many additional variables and constraints will be introduced into the FCOPF framework when incorporating the DL-based frequency predictor, and it also determines the accuracy of the original DL model, thereby influencing the overall effectiveness of the DL-FCOPF. Therefore, sensitivity analysis is conducted in this paper to identify the most suitable DL model.

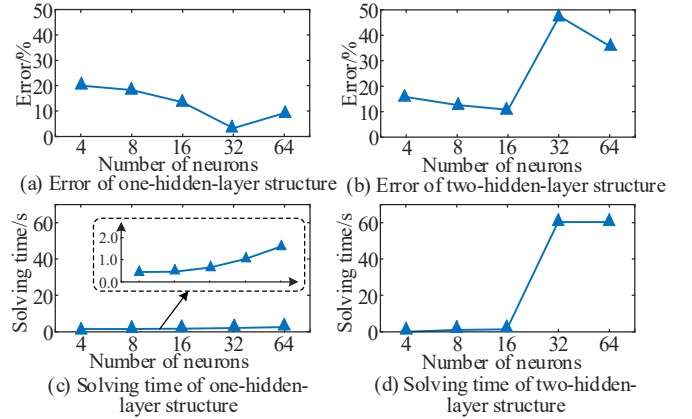

Fig. 10. Comparisons of results between DL-FCOPF for different DL model structures: error of a) one-hidden-layer structure, b) two-hidden-layer structure, and solving time of c) one-hidden-layer structure, d) two-hidden-layer structure.

Fig. 10 presents a comparison of the error and solving time of the DL-FCOPF for different DL model structures. Both one-hidden-layer and two-hidden-layer DL models with varying neuron counts are examined. All models use the ReLU as the activation function, and the exact linearization approach

BPWL is applied to eliminate the linearization error and accurately assess the influence of the number of hidden layers and neurons. The error in Fig. 10 is the mean square error of the four features: RoCoF, FN, headroom of IBRs at bus 36 and bus 39 between the DL-FCOPF results and EMT simulation results. The solving time represents the computation time required to solve the DL-FCOPF.

Fig. 10 (a) shows that for the one-hidden-layer structure, as the number of neurons increases, the error of the DL-FCOPF results decreases. When the number of neurons reaches 32, the error attains its minimum value. Beyond this point, the error increases again due to potential overfitting of the model. Fig. 10 (c) illustrates the solving time for the one-hidden-layer DL-FCOPF. Although the number of neurons increases, the solving time does not rise significantly and can be further reduced by applying different linearization approaches to the activation function, which will be discussed in Section V.B.2. For the two-hidden-layer structure in Fig. 10 (b) and (d), the first hidden layer contains 32 neurons, while the number of neurons in the second hidden layer varies from 4 to 64. As shown in Fig. 10 (b), when the number of neurons in the second hidden layer reaches 32, the error increases sharply. This occurs because the solving time of the DL-FCOPF model with a two-hidden-layer DL-based frequency predictor model reaches the pre-specified 60-second time limit, which is reasonable as OPF is a real-time application and needs to be solved within a very limited time, as illustrated in Fig. 10 (d). Comparing the four subfigures in Fig. 10, the one-hidden-layer DL-FCOPF with 32 neurons achieves the lowest error while maintaining an acceptable solving time. Thus, it is identified as the most suitable model for the proposed frequency predictor in this paper.

*2) Activation function linearization approaches:* Comparing the solving time in Tables II and V, it can be observed that the solving time increases from 0.55s to 1.03s when the system size expands from the 9-bus system to the 39-bus system, which is relatively high compared to a solving time of around 0.2s for T-OPF and L-FCOPF. Therefore, to reduce the solving time while maintaining the accuracy of the DL-FCOPF framework, three approximate linearization approaches illustrated in Fig. 4 are applied to the DL-based frequency predictor and incorporated into the FCOPF framework.

The resulting solutions are compared in Table VII. Since penalty costs are included in P-CTAR and PCAR linearized DL-FCOPF models, the *real cost* in this table refers to the total cost excluding the penalty terms. The *linearization error* stands for the mean square error of four kinds of features: RoCoF, FN, the headroom of GFM IBRs at buses 36 and 39 obtained from the DL-FCOPF framework compared with the original DL model, while the *real error* represents the corresponding differences between the DL-FCOPF results and the EMT simulation results obtained in PSCAD.

TABLE VII
SOLUTIONS OF DL-FCOPF WITH FOUR DIFFERENT LINEARIZATION APPROACHES UNDER SCENARIO 2

| Linearization method | Real cost/$ | $\alpha_{36}$ | $\alpha_{39}$ | Solving time/s | Linearization error | Real error |
|---|---|---|---|---|---|---|
| BPWL | 354518 | 0.49 | 0.48 | 1.03 | 0.00% | 3.18% |
| CTAR | 345171 | 0.16 | 0.18 | 0.42 | 68.43% | 72.44% |
| P-CTAR | 357818 | 0.61 | 0.58 | 0.44 | 0.01% | 3.19% |
| PCAR | 357818 | 0.61 | 0.58 | 0.43 | 0.01% | 3.19% |

Table VII shows that CTAR, P-CTAR and PCAR significantly reduce the solving time by about 60%, as they avoid the introduction of binary variables. Among the three approaches, CTAR yields a high real error at 72.4%, which is unacceptable. Its linearization error reaches 68.42%, indicating that most of the real error originates from the linearization itself. For P-CTAR and PCAR, the real errors mainly stem from the original DL model rather than the linearization process. In contrast, P-CTAR and PCAR achieve low real error of 3.19%, comparable to that of BPWL, while exhibiting shorter solving time. However, the real costs of P-CTAR and PCAR are higher than that of BPWL, as these approaches require larger allocations of GFM and GFL IBRs and thereby greater headroom, resulting in higher real cost.

The penalty coefficient influences the result of P-CTAR and PCAR linearized DL-FCOPF framework. Therefore, different penalty coefficients are applied to this model and then their performance is compared in Fig. 11.

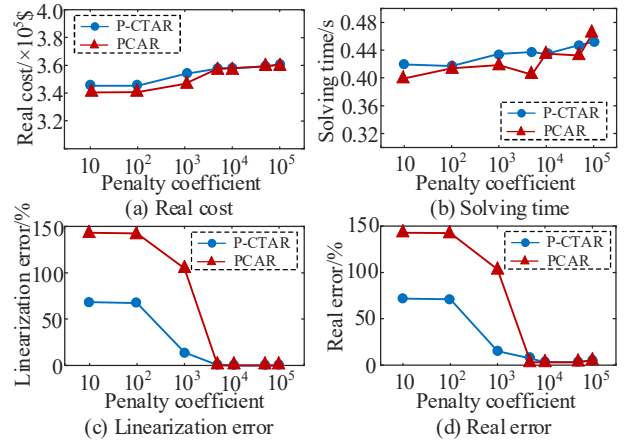

Fig. 11. Comparison of solutions under different penalty coefficients for P-CTAR and PCAR based DL-FCOPF: a) real cost, b) solving time, c) linearization error, d) real error.

Fig. 11(a) shows that the real costs of the P-CTAR and PCAR linearized DL-FCOPF frameworks become identical when the penalty coefficient reaches 5000. In Fig. 11(b), although solving time slightly increases with the penalty coefficient, it remains around 0.4s and does not rise significantly. Fig. 11(c) and (d) indicate that when the penalty coefficient is below 5000, the error between the DL-FCOPF results and EMT simulation results increase noticeably for both P-CTAR and PCAR, primarily due to the linearization approaches. Therefore, when the penalty coefficient exceeds 5000, the P-CTAR and PCAR linearized DL-FCOPF frameworks exhibit similar performance, achieving low errors and acceptable solving time, and can thus be effectively used for DL model reformulation in this paper.

In summary, both the P-CTAR and PCAR provide acceptable ReLU linearization accuracy when the penalty coefficient exceeds 5000.

## VI. CONCLUSIONS

In this paper, a DL-FCOPF framework is proposed to ensure frequency stability while considering optimal allocation of GFM and GFL IBRs and the required headroom for allocated GFM IBRs. The DL-based frequency predictor for frequen-cy-dynamics-related metrics is first introduced, and then reformulated and integrated into FCOPF to establish the proposed DL-FCOPF. For comparison, the T-OPF and L-FCOPF are also developed as two benchmark models to evaluate the effectiveness and efficiency of the proposed framework.

Case studies conducted on the modified WSCC 9-bus and IEEE 39-bus systems indicate that the proposed DL-FCOPF has a lower error of frequency dynamics comparing with simulation results in PSCAD, while T-OPF lacks that information and L-FCOPF has a larger error. Furthermore, ignoring the allocation of GFM and GFL IBRs and their frequency support capability may cause insufficient active power support during contingencies or lead to the economic inefficiency of the power system. Finally, sensitivity analyses indicate that the most suitable configuration of the DL-based frequency predictor is one hidden-layer using either the P-CTAR or PCAR linearization approach.

ACKNOWLEDGMENT

This research was in part supported by the National Science Foundation (NSF) under Grant No. 2337598 and the University of Houston Chevron Energy Graduate Fellows Awards.